\documentclass[apj]{emulateapj}

\usepackage{apjfonts}

\newcommand{\twomass}{2MASS}
\newcommand{\akari}{{\it AKARI}}
\newcommand{\apex}{APEX}
\newcommand{\alma}{Atacama Large Millimeter/sub-millimeter Array}
\newcommand{\gs}{Gemini S}
\newcommand{\herschel}{{\it Herschel}}
\newcommand{\hipp}{{\it Hipparcos}}
\newcommand{\iras}{{\it IRAS}}
\newcommand{\iso}{{\it ISO}}
\newcommand{\spitzer}{{\it Spitzer}}
\newcommand{\wise}{{\it WISE}}

\newcommand{\fis}{FIS}
\newcommand{\gpi}{Gemini Planet Imager}
\newcommand{\irs}{IRS}
\newcommand{\laboca}{LABOCA}
\newcommand{\mips}{MIPS}

\newcommand{\sphere}{Spectro-Polarimetric High-contrast Exoplanet Research}
\newcommand{\trecs}{T-ReCS}

\newcommand{\um}{$\mu$m}

\newcommand{\bem}{$\beta$}

\newcommand{\lam}{$\lambda$}

\newcommand{\dgr}{$^{\circ}$}

\newcommand{\hdtar}{HD 131835}
\newcommand{\hdstd}{HD 136422}

\begin{document}
\submitted{Accepted for Publication in ApJ}
\title{Discovery of Resolved Debris Disk Around HD 131835}
\shorttitle{Resolved Debris Disk Around HD 131835}
\shortauthors{HUNG et al.}
\author{Li-Wei Hung\altaffilmark{1}, Michael P. Fitzgerald\altaffilmark{1}, Christine H. Chen\altaffilmark{2}, Tushar Mittal\altaffilmark{3}, Paul G. Kalas\altaffilmark{4}, and James R. Graham\altaffilmark{4}}

\altaffiltext{1}{Department of Physics and Astronomy, University of California,
Los Angeles, CA 90095, USA; liweih@astro.ucla.edu, mpfitz@ucla.edu}
\altaffiltext{2}{Space Telescope Science Institute, 3700 San Martin Drive, Baltimore, MD 21218, USA}
\altaffiltext{3}{Department of Earth and Planetary Science, University of California, Berkeley, CA 94720, USA}
\altaffiltext{4}{Department of Astronomy, University of California, Berkeley, CA 94720, USA}

\begin{abstract}

We report the discovery of the resolved disk around \hdtar\ and present the analysis and modeling of its thermal emission. \hdtar\ is a $\sim$15 Myr A2 star in the Scorpius-Centaurus OB association at a distance of 122.7$^{+16.2}_{-12.8}$ parsec. The extended disk has been detected to $\sim 1.5\arcsec$ (200 AU) at 11.7 \um\ and 18.3 \um\ with \trecs\ on Gemini South. The disk is inclined at an angle of $\sim$75\dgr\ with the position angle of $\sim$61\dgr. The flux of \hdtar\ system is $49.3 \pm 7.6$ mJy and $84 \pm 45$ mJy at 11.7 \um\ and 18.3 \um\ respectively. A model with three grain populations gives a satisfactory fit to both the spectral energy distribution and the images simultaneously. This best-fit model is composed of a hot continuous power-law disk and two rings. We characterized the grain temperature profile and found that the grains in all three populations are emitting at temperatures higher than blackbodies. In particular, the grains in the continuous disk are unusually warm; even when considering small graphite particles as the composition.

\end{abstract}



\section{INTRODUCTION}

Planet formation and evolutionary history is imprinted on the nature and distribution of circumstellar debris. Circumstellar debris disks are composed of dust and planetesimals that are in orbit around main sequence stars, analogous to the asteroid belt and Kuiper belt in our solar system. The materials in the disks are thought to be the second generation where the dust is constantly being replenished \citep{back93} from collisions between planetesimals and sublimation of comets \citep{harp84}. Most of the debris disks are discovered through detecting their infrared excess and characterized by their spectral energy distributions (SEDs). However, studying their SEDs alone provides only limited information. While the disk temperature and the total infrared (IR) flux can be well determined from the SEDs, the spatial information of the systems is inaccessible. In addition, there is a possibility that the excess actually originates from an unrelated foreground or background source contaminating the beam.

In addition, with the SED information only, the location of the grains will be degenerate with the grain properties. For example, small grains further away from the star might be heated to the same temperature as large grains that are closer toward the center. Thus, a single-temperature debris disk does not necessarily indicate all the grains are located in a thin ring. A similar argument applies to the opposite scenario where small grains might be heated to a higher temperature compared to large grains at the same location. Therefore, for a multiple-temperature debris disk, the SED information alone is typically insufficient to distinguish between the cases where grains are located at the same spatial location (i.e., a ring) verses at multiple spatially distinct locations (i.e., belts or extended disks). \citet{kenn14} argued that this degeneracy is small for A-type stars, due to the truncation on the small end of the grain size distribution by radiation pressure. They proposed that most two-temperature disks around A-type stars probably arise from multiple spatial components. To confirm that the disk indeed has multiple spatial components, resolved images are required.

If the disks can be spatially resolved, we will be able to characterize the spatial distribution of the grains and the geometry of the disks. We can use distribution of grains to probe the underlying physics responsible for dust distribution, and constrain the locations and nature of the grain parent bodies. In addition, the morphology of the planetesimal belts might be closely related to the presence of unseen planets dynamically sculpting the disk. For example, as seen in mid-IR thermal emission, HR 4796A has an inclined ring with one lobe being $\sim$5\% brighter than the other \citep{tele00}. \citet{wyat99} argued that the secular perturbations from a planet of mass > 0.1 $M_{\rm J}$ located close to the inner edge of the disk could introduce this brightness asymmetry. The planet would impose the forced eccentricity on the ring, causing one side of the ring being closer to the star and thus showing the "pericenter glow." Such incidence demonstrates the level of information provided from the direct imaging far surpasses from the SED alone. 

Here we have attempted to spatially resolve the disk around \hdtar, a system characterized solely by its SED prior to this work. \citet{rizz11} assign a 91\% membership probability to the Upper Centaurus Lupus (UCL) moving group (a subgroup of the Sco-Cen association) based on its Galactic location and velocity. \hdtar\ is a young star ($\sim$15 Myr, based on the age of the UCL estimated by \citealt{mama02}; $\sim$16 Myr, based on an analysis of F-type pre-main-sequence members of the group by \citealt{peca12}) with a spectral type of A2IV \citep{houk82} at the distance of d = 122.7 $^{+16.2}_{-12.8}$ pc \citep{vanl07}. The infrared excess emission of \hdtar\ was first reported by \citet{moor06} through searching the {\it Infrared Astronomical Satellite} (\iras) and {\it Infrared Space Observatory} (\iso) databases for the stars in the vicinity of the Sun. The \mips\ data for this source were first published in \citet{chen12} which shows the \mips\ 24 and 70 \um\ data for all of the \citet{deze99} ScoCen A-type stars and helps to put the excess emission for this star into context (e.g. it is one of only 4 UCL/Lower Centaurus Crux A-type stars with $\rm L_{\rm IR} / L_\ast$ commensurate to beta Pic, $> 10^{-3}$). \citet{chen14} found that the SED for \hdtar\ can be described using a two-temperature model.

In contradiction to the traditional disk evolution scheme, CO gas in debris disks around A-type stars may be fairly common. Typically, the primordial gas dissipates in a few million year time scale during the protoplanetary phase. By the time it transforms into a debris system, the CO gas level will be undetectable. However, there are several exceptional cases. For example, 49 Cet and HD 21997 are 40 \citep{zuck12} and 30 \citep{torr08} Myr old A-type stars hosting gas-rich disks (\citealt{zuck95}; \citealt{moor11}). The gas is believed to be the second generation, possibly due to violent comet collisions \citep{zuck12}. In the case of HD 21997, there is an alternative explanation based on a recent ALMA observation which indicates that the gas may be of primordial origin in this system \citep{kosp13}. Another possible production mechanism involves constant resurfacing of the parent bodies and sublimation or photodesorption of the CO ice. For the $\beta$ Pic system, its CO distribution is particularly interesting because it is not the same as the dust, even more highly asymmetric and may imply the presence of a second undetected planet \citep{dent14}. Not only has CO been detected and characterized for $\beta$ Pic, 49 Cet, and HD 21997 but it has also been detected around 5 A-type stars in Upper Sco (Hughes 2014, private communication).
 
Besides having the IR excess from the dusty debris, \hdtar\ hosts a detectable amount of carbon monoxide gas (\citealt{moor13}, Mo{\'o}r et al. in prep). For \hdtar, \citet{kast10} first reported the nondetection of CO emission with the 30-m Institut de Radio Astronomie Millimetrique telescope. However, \citet{zuck12} argued that if the comet collision model is correct, then the H$_2$/CO ratio is unconstrained and thus the upper mass limit of the CO gas for \hdtar\ is $4.06 \times 10^{-3}$ M$_E$. Later, \citet{moor13} announced the discovery of detected submillimeter CO emission with the Atacama Pathfinder EXperiment (APEX) radio telescope through a survey. Once the CO gas is well characterized, it can provide us valuable information on the disk environment and the dust-gas associations of such system with the coexistence of gas and debris at an age $\gtrsim$10 Myr.

Here we report the discovery of the spatially resolved debris disk around \hdtar\ at mid IR wavelengths, tracing the thermal emission from the grains. Currently, these images are the only resolved images of the system. In this paper, we present the analysis, modeling, and characterization of the debris disk around \hdtar. Our mid IR images are based on two epochs of observation as described in \S~\ref{obs}. The data processing details, including image reduction, photometry, and PSF subtraction, are in \S~\ref{dataprocess}. In \S~\ref{analysis}, we analyze the stellar properties and show that a three-component model is required to describe the system. In \S~\ref{discussion}, we characterize the grain temperatures in the system and investigate the possible grain compositions. We also briefly discuss how the current generation of imaging instruments and telescopes can improve our understanding of debris disks in the near future. Finally, we summarize our modeling results in \S~\ref{summary}.

\section{OBSERVATIONS}
\label{obs}

We used the \trecs\ instrument on the Gemini South telescope to obtain mid-IR imaging of HD 131835 in two programs, GS-2008A-Q-40 and GS-2009A-Q-19. As part of these programs, the star \hdstd\ was also observed as a photometric and point-spread function (PSF) calibration standard. The observations of the reference star were interleaved between the observations of the target. All the images are taken in the Si-5 (\lam$_c$ = 11.66 \um, $\Delta$\lam = 1.13 \um) and Qa (\lam$_c$ = 18.30 \um, $\Delta$\lam = 1.51 \um) filters. In these bands, the \trecs\ pixel size is $0.08976\arcsec \pm 0.00021\arcsec$ on the sky.\footnotemark\ Due to the high sky background and instrumental thermal background, we used the standard mid-IR ABBA chop-and-nod observing technique with a 15$\arcsec$ throw between chop-nod positions. 

\footnotetext{http://www.gemini.edu/sciops/instruments/t-recs/spectroscopy/detector}

Since the disk had not been resolved prior to the 2008 observations, the first epoch data were obtained with an arbitrary chop position angle. This turned out to be very close to the apparent position angle of the disk. To confirm that the feature was truly from the emission of the resolved disk and not from artifacts due to imperfect chopping and nodding, the second epoch of observation was obtained in 2009 with the chop position angle roughly perpendicular to the previous one. Observation conditions were generally very good, with diffraction rings visible in the calibration images. The sky was very transparent, with low water vapor (<1mm) measured towards zenith for the most of the nights. The processed images of HD 136422 gave diffraction-limited resolution of $\sim 0.39\arcsec$ and $\sim 0.54\arcsec$ at 11.7 \um\ and 18.3 \um\ respectively. \hdtar\ (and \hdstd) were observed in two nights in 2008 and four nights in 2009, with the total on source time of 6168 (2056) seconds at 11.7 \um\ and 5734 (1303) seconds at 18.3 \um. Details of the observations are summarized in Table~\ref{observations}.

\begin{deluxetable}{cccccc}
\tablecaption{Observations with Gemini S / T-ReCS \label{observations}}
\tablewidth{0.in}
\tablehead{
\colhead{Program ID}&
\colhead{PA}&
\colhead{UT Date}&
\colhead{Star}&
\colhead{Filter}&
\colhead{Integration} \\
\colhead{(GS-20+)}&
\colhead{(deg)}&
\colhead{}&
\colhead{}&
\colhead{}&
\colhead{Time (sec)}
}
\startdata
08A-Q-40 & 55 & 2008-05-08 &\hdtar &Si-5 & \ 318.5\ \ \ \ \ \ \\
&&&&Qa&1216.3\ \ \ \ \ \ \\
&&&\hdstd&Si-5& \ \ \ 86.9\ \ \ \ \ \ \\
&&&&Qa& \ 144.8\ \ \ \ \ \ \\
&& 2008-05-11 &\hdtar &Si-5 & \ 637.1\ \ \ \ \ \ \\
&&&\hdstd&Si-5&\ \ \ 86.9\ \ \ \ \ \ \\
09A-Q-19 & 148 & 2009-04-18 &\hdtar & Qa &3011.7\ \ \ \ \ \ \\
&&&\hdstd&Qa&\ 724.0\ \ \ \ \ \ \\
&& 2009-04-19 &\hdtar &Si-5 &3127.5\ \ \ \ \ \ \\
&&&\hdstd&Si-5&1013.5\ \ \ \ \ \ \\
&& 2009-05-23 &\hdtar &Si-5 &1042.5\ \ \ \ \ \ \\
&&&&Qa&1505.8\ \ \ \ \ \ \\
&&&\hdstd&Si-5&\ 434.4\ \ \ \ \ \ \\
&&&&Qa&\ 434.4\ \ \ \ \ \ \\
&& 2009-07-12 &\hdtar &Si-5 &1042.5\ \ \ \ \ \ \\
&&&\hdstd&Si-5&\ 434.4\ \ \ \ \ \ \\
\cutinhead{Total}
&&&&\\
&&&\hdtar&Si-5& 6168.2\ \ \ \ \ \ \\
&&&&Qa& 5733.8\ \ \ \ \ \ \\
&&&\hdstd&Si-5& 2056.0\ \ \ \ \ \ \\
&&&&Qa& 1303.1\ \ \ \ \ \ 
\enddata
\end{deluxetable}

\section{DATA PROCESSING}
\label{dataprocess}

\subsection{Image Reduction}
\label{imgreduction}

\begin{figure*}
\begin{center}
\includegraphics[angle=0,width=6in]{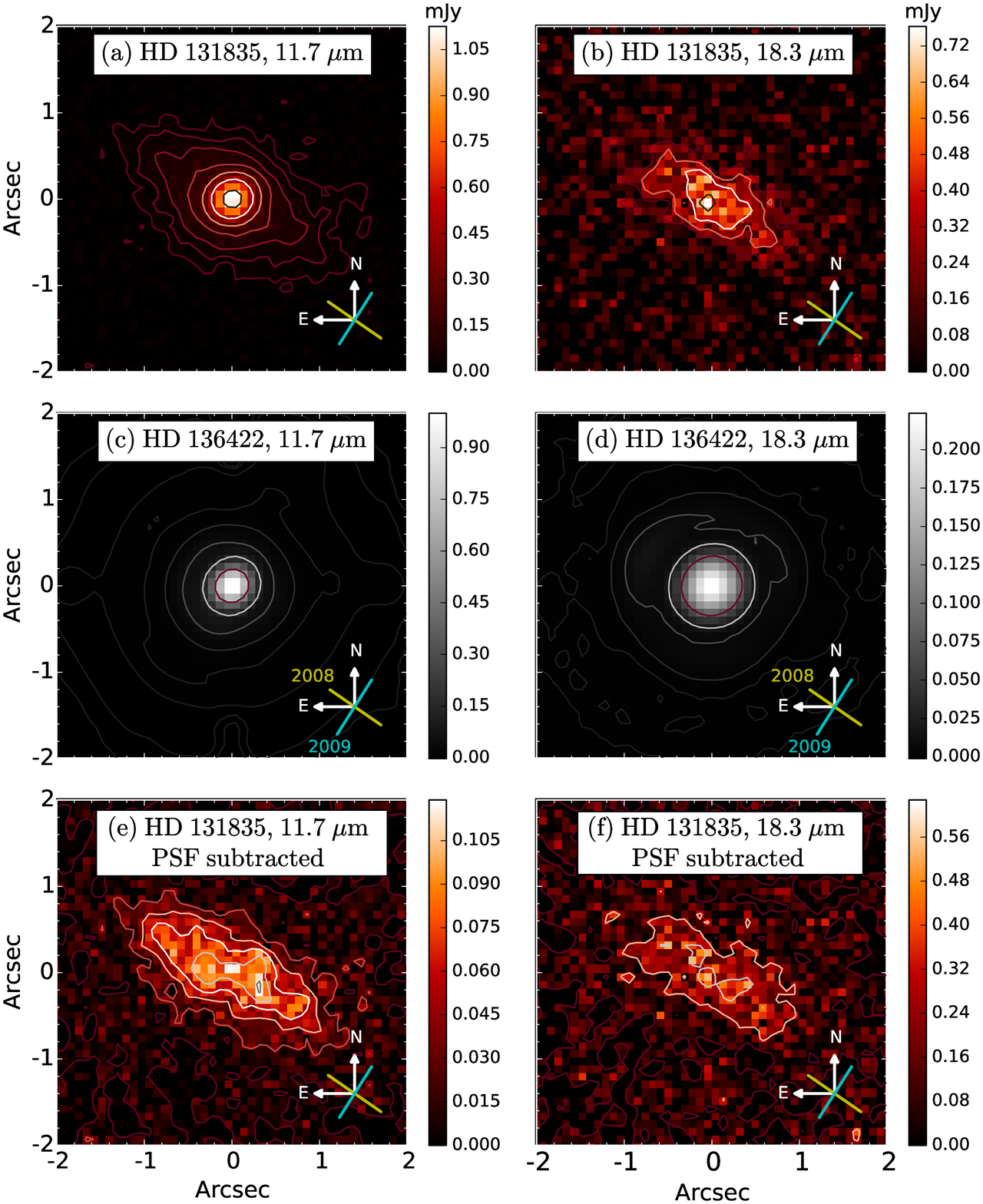}
\end{center}
\setlength{\abovecaptionskip}{-10pt}
\caption{(a) and (b) are the images of \hdtar\ from \trecs\ on Gemini South at 11.7 \um\ and at 18.3 \um. (c) and (d) are the corresponding images of the reference star \hdstd\ scaled to the stellar flux of the target. Contours in (a)-(d) are spaced in logarithmic scales. (e) and (f) are the PSF subtracted images of \hdtar, and contours are spaced by the 1-$\sigma$ background noise level. In both wavelengths, the disk is resolved out to approximately 200 AU.  The lines on the lower right-hand corners indicate the chop position angles for observations in 2008 and 2009 epochs.
\label{psfsubimg}}
\end{figure*}

We reduced all the images following the standard mid-infrared data reduction procedures. We linearly combined the chop and nod frames within each image to remove the instrumental and sky background. We then subtracted the average row background and corrected for the column offset due to small bias variations among different channels. The cores of the PSF standard star \hdstd\ are fairly circular (as shown in Fig.~\ref{psfsubimg} c and d) with the FWHM of $\sim 0.39\arcsec$ and $\sim 0.54\arcsec$ at 11.7 \um\ and 18.3 \um\ respectively. The target images in both wavelengths show the extended and elongated emission compared to the PSFs. Before the PSF subtraction process, these mid-IR images already show prominent disk emission extending beyond $1\arcsec$ along the semi-major axis in both bands.  

\subsection{Photometry}
\label{photometry}

We performed aperture photometry on \hdtar\ using \hdstd\ as the photometric reference. The flux of \hdstd\ was taken from \citet{cohe99}. We applied aperture correction to our photometry measurement on \hdtar. Based on the observations of \hdstd, the PSF structure was fairly stable over the course of each night; the enclosed flux in an aperture radius of 1$\arcsec$ usually varied only from 1 to 3 \%. Thus, we adopted one relationship of aperture radius versus enclosed flux fraction for each night. Due to the nonuniform residual background, the enclosed flux fluctuated at large aperture radii. We characterized these fluctuations as the uncertainties of the radius-enclosed flux relationships.

Next, we applied atmospheric extinction correction and color correction. The atmospheric extinction varied from 0.09 to 0.19 magnitude per airmass among different nights in both bands. These values are consistent with the extinction observed on Mauna Kea \citep{kris87}, where the conditions are similar to Cerro Pach{\'o}n. For color correction, since the 11.7 \um\ and 18.3 \um\ fell within the wavelength coverage of the \irs\ spectrum, we used its shape to compute the correction factors. Detailed information about the IRS data is provided in \S~\ref{analysis}. The final flux of \hdtar\ is $49.3 \pm\ 7.6$ mJy for 11.7 \um\ and $84 \pm\ 45$ mJy for 18.3 \um. 

\begin{figure}
\begin{center}
\hspace*{-0.15in}
\includegraphics[angle=0,width=3.8in]{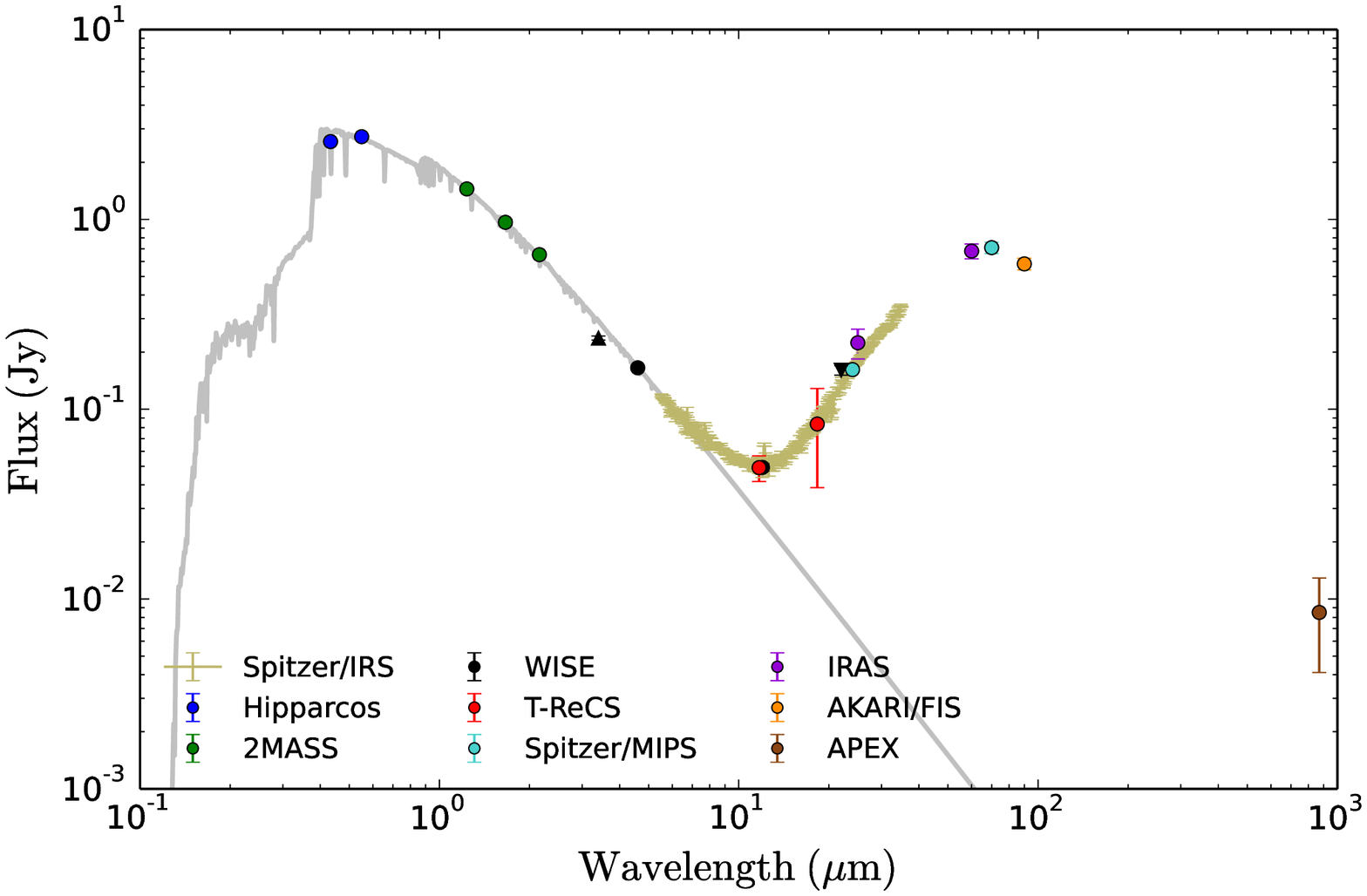}
\end{center}
\setlength{\abovecaptionskip}{-10pt}
\caption{SED of the \hdtar\ system. There are 15 well characterized photometry points and an \irs\ spectrum. The 11.7 \um\ and 18.3 \um\ \trecs\ photometry points are measured in this paper. Reference of other points are listed in Table~\ref{sedtable}. Note that the two triangle \wise\ points in 3.4 \um\ and 22 \um\ are the lower and upper limits. Well characterized photometry points between 3 and 100 \um\ are color corrected. Only the points with wavelength < 5.8 \um\ are used to fit for the stellar atmosphere. The model is described in \S~\ref{steprop}, and the gray line shows the best fit.
\label{sedfig0}}
\end{figure}

We have considered uncertainties from the process of measuring the target's flux, the standard star's flux, the aperture correction, and the extinction correction. However, by comparing the photometry measurement from each image (33 images at 11.7 \um\ and 14 images at 18.3 \um), we noticed the systematic errors dominated over the random errors. Thus, being conservative, we quoted the sample standard deviations of photometry measurements from the ensemble of images as the uncertainties of our mean fluxes. Our measured fluxes are consistent with the values from the \irs\ spectrum. Table~\ref{sedtable} lists the flux measurements of \hdtar\ from this work and from other published literature, and the SED is shown in Fig.~\ref{sedfig0}. 

\begin{deluxetable*}{cccccc}
\tablecaption{Summary of the Measured Fluxes of \hdtar \label{sedtable}}
\tablewidth{7.1in}
\tablehead{
\colhead{}&
\colhead{}&
\colhead{}&
\colhead{Color Corrected}&
\colhead{}\\
\colhead{Wavelength (\um)}&
\colhead{Source}&
\colhead{Flux (mJy)}&
\colhead{Flux (mJy)}&
\colhead{References}}
\startdata
0.43 &\hipp                            &2571 $\pm$ 48                            & &(1)\\
0.55 &\hipp                            &2728 $\pm$ 26                            & &(1)\\
1.24 &\twomass                         &1448 $\pm$ 29                            & &(2)\\
1.66 &\twomass                         &\ \ 965 $\pm$ 35                         & &(2)\\
2.16 &\twomass                         &\ \ 652 $\pm$ 11                         & &(2)\\
3.4  &\wise                            &\ \ \ \ 237.2 $\pm$ 5.8 \tablenotemark{a,b}    & &(3)\\
4.6  &\wise                            &\ \ 166.8 $\pm$ 4.7 \tablenotemark{a}      &165.3 $\pm$ 4.7 \ \ &(3) (4)\\
11.7 &\gs\ / \trecs                    &\ \ 49.7 $\pm$ 7.7                     &49.3 $\pm$ 7.6  &(5)\\
12   &\wise                            &\ \ \ \ 49.1 $\pm$ 2.2 \tablenotemark{a}   &49.4 $\pm$ 2.2 &(3) (4)\\
18.3 &\gs\ / \trecs                    &\ \ \ \ 83 $\pm$ 44                    &\ \ 84 $\pm$ 45    &(5)\\
22   &\wise                            &\ \ \ \ 160.5 $\pm$ 9.4 \tablenotemark{a,c}    & &(3)\\
24   &\spitzer\ / \mips                &153.1 $\pm$ 3.1                          &161.7 $\pm$ 3.3 \ \ &(6) (7) (8)\\
25   &\iras                            &\ \ 186 $\pm$ 34                       &224 $\pm$ 40    &(9)\\
60   &\iras                            &\ \ 684 $\pm$ 62                       &681 $\pm$ 61    &(9)\\
70   &\spitzer\ / \mips                &\ \ \ \ 659.2 $\pm$ 44.7 \tablenotemark{d} &710.0 $\pm$ 48.1 &(7) (8) (10)\\
90   &\akari\ / \fis\ \ \ \            &\ \ 560 $\pm$ 39                         &583 $\pm$ 41 &(11) (12)\\
870  &\ \ \ \ \ \ \ \ \apex\ / \laboca &\ \ \ \ 8.5 $\pm$ 4.4                    & &(13)\\
\cutinhead{Spectrum}
5.2 -- 37.9&\spitzer\ / \irs   & ... &         &(6)
\enddata
\tablenotetext{a}{The quoted WISE data are based on aperture photometry. The error has included the uncertainty from the RMS scatter in the standard calibration stars.}
\tablenotetext{b}{This is a lower limit due to saturation.}
\tablenotetext{c}{This is an upper limit due to source confusion.}
\tablenotetext{d}{The error is calculated by considering the source photon counting uncertainty, the detector repeatability uncertainty, and the absolute calibration uncertainty}
\tablerefs{(1) \citealt{hog00}; (2) \citealt{cutr03}; (3) \citealt{cutr12}; (4) \citealt{wrig10}; (5) this paper; (6) \citealt{chen14}; (7) MIPS Instrument Handbook (Version 3.0, March 2011) by MIPS Instrument and MIPS Instrument Support Teams; (8) \citealt{riek04}; (9) \citealt{moor06}; (10) \citealt{chen12}; (11) \citealt{yama10}; (12) \citealt{liu14}; (13) \citealt{nils10}}
\end{deluxetable*}

\subsection{PSF Subtraction}
\label{psfsub}

After linearly combined the ABBA frames within each fits image, we weighted combined all these individual image files to produce one final high signal-to-noise ratio (SNR) image in each band. The uncertainty for each pixel in the final images was calculated by propagating the errors through the combination process. The PSF was constructed by a similar process except for the weight used for stacking the target images.

The target, PSF, and PSF-subtracted final images are shown in Fig.~\ref{psfsubimg}. The diagonal lines in the figure represent the chop position angles. The chop position angle was rotated roughly $90^{\circ}$ from 2008 to 2009 due to the resolved disk structure as described in \S~\ref{obs}. We scale the PSF to the target star flux, then subtract the stellar flux out from the target images. \S~\ref{steprop} describes  the stellar flux characterization in detail. By processing the 2008 and 2009 data separately, we see that 2009 observation alone confirms the detection of the structure and shows the position angle of the disk that is consistent with the observations made in 2008. Thus, we are confident that the elongated structure detected in 2008 is real, rather than an artifact arising due to imperfect chop-nod motion.

In both bands, the central region of the disk is generally brighter than the outskirt. The detected SNR is greater than 6 and 2 at 11.7 \um\ and 18.3 \um\ respectively. At 11.7 \um\, the extended structure is detected at 1 $\sigma$ out to $\sim 1.5\arcsec$ and 0.7$\arcsec$ in the major and minor axes, corresponding to $\sim 180$ AU and $\sim 85$ AU from the star in projection. The 18.3 \um\ emission is resolved out to $\sim 130$ AU. Such extended disk structure in the mid-IR suggests the presence of warm dust far away from the star.

\section{MODELING AND ANALYSIS}
\label{analysis}
The infrared excess of \hdtar\ has been measured in multiple wavelengths. Table~\ref{sedtable} shows the summary of the measurements taken from the published literature. \hdtar\ was detected in all four \wise\ bands \citep{cutr12}. However, the \wise\ w1 measurement represents a lower limit due to saturation, and w4 measurement is an upper limit due to source confusion, as flagged in the \wise\ catalog by \citet{cutr12}. The \iras\ photometry measurements are taken from \citet{moor06}. The \spitzer/\mips\ 24 \um\ and 70 \um\ measurements are taken from \citet{chen14} and \citet{chen12}. The object 1456545-354138 in the \akari\ catalog \citep{yama10} was identified as \hdtar\ by \citet{liu14}. Among the four \akari/\fis\ bands, only the source measured in 90 \um\ is confirmed and reliable \citep{yama10}. Finally, the longest wavelength point on the SED was measured in 870 \um\ by \citet{nils10}.

We apply color correction to good photometry measurements between 3 \um\ and 100 \um. \citet{moor06} had computed the color corrected fluxes for the \iras\ points. For other points that fall within the wavelength coverage of the \irs\ spectrum, we use the shape of the \irs\ spectrum to determine the correction factors. The other photometry points are corrected based on the best estimated local SED shape. Color-corrected fluxes and references used to compute the correction factors are listed in Table~\ref{sedtable}. We do not consider correcting the 870 \um\ point since it is on the Rayleigh-Jeans tail so the correction factor should be close to unity. If a point is color-corrected, the corrected flux is used in the following modeling and analysis. Fig.~\ref{sedfig0} shows the SED (with the color-corrected fluxes). The SED shows a single hump infrared excess with the flux peaking around 70 \um. The system has $\rm L_{\rm IR} / L_\ast\ \sim 2 \times 10^{-3}$.  

The \spitzer\ \irs\ data were obtained and analyzed by \citet{chen14}. It is generally presumed that the \mips\ fluxes will have a lower absolute calibration uncertainty. Therefore, \citet{chen14} calibrated the \irs\ spectrum by tabulating the synthetic \mips\ flux and comparing it to the measured \mips\ flux. They presumed that the source did not vary with time and then scaled the \irs\ spectrum to match the \mips\ 24 \um\ flux. More details of the spectral extraction and calibration can be found in \citet{chen14}. The calibrated \irs\ spectra do not show any obvious solid state emission or absorption features, suggesting \hdtar\ is not a system rich in small silicate grains and polycyclic aromatic hydrocarbons. For the following analysis, we treat each point of the spectra as a single measurement that is directly comparable to photometry measurements. When including images in the fit, each pixel is considered as a data point and is treated equally as a point on the SED. For each image, only the central 68 by 68 pixels ($\sim 6 \arcsec$ by  $6 \arcsec$) are considered in the fit. There are total of 9595 data points in which 10 points come from broadband photometry, 337 points come from the \irs\ spectrum, and 9248 points from the images.

\subsection{Stellar Properties}
\label{steprop}
The measured flux of the system includes the contribution from the star and from the disk. The star has $T_{\rm eff}$ = 8770 K, log$g$ = 4.0 and solar metallicity with A(V) = 0.187 mag \citep{chen12}. We use the IDL Astrolib routine ccm\_unred.pro, which is based on \citet{card89}, to apply reddening to the \citet{kuru93} stellar atmosphere model. Since the disk emission is prominent in longer wavelengths, only the measurements with wavelength shorter than 5.8 \um\ are used to fit for the stellar flux to characterize the stellar contribution. We compare the \hipp\ and \twomass\ photometry to the synthetic fluxes of the model during the fitting process. The best fit is shown in Fig.~\ref{sedfig0}. The stellar fluxes at 11.7 \um\ and 18.3 \um\ of this best-fit model were used to scale the PSFs during the PSF subtraction process described in \S~\ref{psfsub}. 

We parameterize a scaling factor as $\xi\ \tbond\ (R_\ast/d)^2$ so that we can write
\begin{equation}
L_\ast\ = 4 \pi d^2 \xi \sigma T^4 = 4 \pi d^2\int_0^\infty I_\nu\ d\nu.
\end{equation}
By integrating the unreddened stellar component of the best-fit SED model and solving the equation, we obtain $\xi = 6.68 \times 10^{-20}$. When estimating the luminosity, the largest uncertainty comes from the distance measurement. Taking the distance to be $122.7\pm^{16.2}_{12.8}$ pc \citep{vanl07}, we find $R_\ast = 1.41\pm^{0.19}_{0.15} R_\sun$ and $L_\ast = 10.5\pm^{2.8}_{2.2} L_\sun$.

\subsection{Model}
\label{model}

Our goal is to find a disk model that can reproduce the observed SED and the images simultaneously. The SED sets strong constraints on the grain temperature distribution, where as the images inform the spatial distribution of the grains. We aim to recover the extended mid-IR emission, with the disk flux peaking close to the center of the system, while simultaneously reproducing the broad IR excess hump on the SED. With the space and temperature distributions of the grains so constrained, we hope to further characterize the grain properties such as its size and composition. In our model, we assume the disk to be optically thin since its $\rm L_{\rm IR} / L_\ast\ \sim 2 \times 10^{-3}$. To search for a model that accounts for the extended emission, we start with a simple two-dimensional continuous disk composed of only a single population of grains, then consider more complicated distributions.

\subsubsection{A Continuous Power-law Disk Model}
\label{model1}

In contrast to perfect blackbodies, dust grains do not couple efficiently to radiation when the wavelength is much larger than the grain size. We consider a simple parameterization of the emissivity with a modified blackbody function, whereby the emissivity varies with the frequency as a power-law with a positive index \bem. The grain temperature $T_g$ at a distance $r$ away from the star is found by balancing the energy intake and the energy output:
\begin{equation}
\label{tgeqn}
\frac{d^2}{r^2} \int F_{\nu, \ast}\ \nu^\beta\ d\nu\ =\ 4 \pi \int B_\nu[T_g(r)]\ \nu^\beta d\nu
\end{equation}
$B_\nu(T)$ here is the blackbody function. The grains are distributed with the surface density of $n_0\ (r/1AU)^\Gamma$, with $n_0$ being the two dimensional number density of the grains at 1 AU. The density law applies to the region between the inner disk radius $r_i$ and the outer radius $r_o$; it is zero elsewhere. The total disk flux $F_{\nu,\rm disk}$ is
\begin{equation}
F_{\nu, \rm disk}\ =\ C\ \int_{r_i}^{r_o} B_\nu[T_g(r)]\ \left(\frac{\nu}{\rm 1\ Hz}\right)^\beta\ \left(\frac{r}{\rm 1\ AU}\right)^{\Gamma+1} dr
\end{equation}
$C$ is the scaling factor such that $C = 2 \pi^2 n_0 a_0^2 {\rm (1 AU)} d^{-2}$. The characteristic grain radius is denoted as $a_0$. 

There are seven free parameters total in this model. Five of which are needed to calculate the total disk flux: \bem, $\Gamma$, $C$, $r_i$, and $r_o$. Two additional parameters, inclination $i$ and position angle $\phi$, are needed to generate model images. We took several steps to initialize the starting parameters for the fit. First, we set \bem\ to zero as if the emission came from perfect blackbodies. Then, we set $r_i$ and $r_o$ to a range of a few pixels to several hundred AU. Afterwards, values for $\Gamma$ and $C$ were obtained by fitting the model to the SED data only. Finally, $i$ and $\phi$ were initialized to the best-fit inclination and position angle when fitting a ring to the images only. Once the parameters are initialized, we used a Levenberg-Marquardt least-squares fitter. The fitting process tends to drive $r_o$ to unbounded values. The reason that $r_o$ tends to drift to unbounded values is due to the model's under-prediction of the flux beyond 35 \um. During the fitting process, after the model settles with the parameters that fit the majority of the data, the model will try to extend $r_o$ to an arbitrary large value so that it will contain slightly more cold dust. The contribution from this cold dust does make the SED fit better but the amount is negligible. Thus, we fix $r_o$ to 400 AU where beyond which the image fit does not get better. The best-fit parameters along with the total chi-square ($\chi^2$) and the reduced chi-square ($\chi^2_{\nu}$) are listed in Table~\ref{fitparams}. The contributions to the total chi-square from the broad band photometry, IRS spectrum, and images are 3\%, 31\%, and 66\%.

\begin{deluxetable*}{ccccc}
\tabletypesize{\footnotesize}
\tablecaption{One-, Two-, and Three-population Model Fits \label{fitparams}}
\tablewidth{7.1in} 
\tablehead{
\colhead{Parameter}&
\colhead{A Continuous Disk} &
\colhead{A Continuous Disk + A Ring} &
\colhead{A Continuous Disk + Two Rings} &
\colhead{Unit}}
\startdata									
$\chi^2$	&$	9920	$&$	6255	$&$	5958	$&		\\
$\chi^2_{\nu}$	&$	1.03	$&$	0.65	$&$	0.62	$&		\\
\cutinhead{Continuous Component}	&$		$&$		$&$		$&		\\
$\beta$	&$	0.76	$&$	1.53^{+0.02}_{-0.01}	$&$	1.64\pm0.02	$&		\\
$C$	&$	3.7\times10^{-31}	$&$	(5\pm1)\times10^{-42}	$&$	(9\pm2)\times10^{-44}	$&	[AU$^{-1}$]	\\
$\Gamma$	&$	1.0	$&$	0.48\pm0.05	$&$	0.53^{+0.06}_{-0.07}	$&		\\
$i$	&$	49	$&$	73.8\pm1.0	$&$	74.5^{+0.9}_{-1.0}	$&	[deg]	\\
$\phi$	&$	74	$&$	62.0^{+1.0}_{-1.1}	$&$	61.2^{+1.0}_{-0.9}	$&	[deg]	\\
$r_i$	&$	0.24	$&$	37\pm3	$&$	35\pm3	$&	[AU]	\\
$r_o$	&$	400	$&$	400	$&$	310^{+30}_{-20}	$&	[AU]	\\
\cutinhead{Ring Component I}	&$		$&$		$&$		$&		\\
$\beta$	&		&$	0.27^{+0.05}_{-0.04}	$&$	0.59\pm0.02	$&		\\
$C_r$	&		&$	(7.4\pm1.5)\times10^{-15}	$&$	(5.9^{+0.8}_{-0.7})\times10^{-16}	$&		\\
$r$	&		&$	61^{+8}_{-6}	$&$	105\pm5	$&	[AU]	\\
\cutinhead{Ring Component II}	&		&$		$&$		$&		\\
$\beta$	&		&		&$	0.32\pm0.06	$&		\\
$C_r$	&		&		&$	(4\pm1)\times10^{-14}	$&		\\
$r$	&		&		&$	220\pm40	$&	[AU]	
\enddata									
\tablecomments{The chi-square ($\chi^2$) and the reduced chi-square ($\chi^2_{\nu}$) are the minimum values estimated by least squares fitting. The one-component parameters are from the least squares fitting; uncertainties are not estimated since the model does not describe this system well at all. The two- and three-component model parameters and uncertainties are quoted from the marginal distributions of MCMC results. For the two-component model, the parameters (from top to bottom) correspond to the quoted least squares are ($\beta$, $C$, $\Gamma$, $i$, $\phi$, $ri$, $ro$, $\beta$, $C_r$, $r$) = (1.53, $5.0 \times 10^{-42}$, 0.47, 74.0, 61.8, 37.5, 400, 0.27, $7.2 \times 10^{-15}$, 62)  For the three-component model, the parameters correspond (from top to bottom) to the quoted least squares are ($\beta$, $C$, $\Gamma$, $i$, $\phi$, $ri$, $ro$, $\beta$, $C_r$, $r$, $\beta$, $C_r$, $r$) = (1.64, $1.1 \times 10^{-43}$, 0.48, 74.3, 61.1, 36, 316, 0.59,  $6.2 \times 10^{-16}$, 105, 0.33,  $4.3 \times 10^{-14}$, 232).}
\end{deluxetable*}

This one-component model does not produce a good fit. The modeled flux is too concentrated in the central region of the disk as the residual images show strongly negative values in center due to over subtraction. As a result, the best-fit $i$ and $\phi$ are not trustworthy. In addition, modeled SED under predicts the flux shorter than 16 \um\ and beyond 33 \um. The drive to better fit the short-wavelength part of the SED is inconsistent with the spatial location of flux in the images. This suggests that the power-law spatial distribution is inconsistent with the resolved outer disk flux and the contribution to the warm SED flux. Since this model clearly does not fit the data, we did not investigate further for estimating the uncertainties of the best-fit parameters. This inconsistency could be potentially resolved by considering a multi-component model. We therefore move on to a more sophisticated disk model by adding a second component.

\subsubsection{A Continuous Power-law Disk + A Ring}
\label{model2}

This two grain populations model is composed of a continuous disk and a thin ring. The first component is as described in the section above. The second component assumes a single population of grains located in a narrow ring at a single radius $r$ away from the star. We adopt the same emissivity law in the form of $\nu^{\beta}$ for the ring component. However, this $\beta$ parameter could have a different value than in the continuous component since the optical properties might be different for different grain populations. The flux from a ring model is:
\begin{equation}
F_{\nu, ring}(r)\ =\ C_r\ B_\nu[T_g(r)]\ \left(\frac{\nu}{\rm 1\ Hz}\right)^\beta
\end{equation}
The grain temperature $T_g$ is found by solving the Equation~\ref{tgeqn}, and $C_r$ is a scaling factor for the ring model. There are five parameters in the ring component: \bem, $C_r$, $i$, $\phi$, and $r$. Limited by the image resolution, we make the continuous disk and the ring sharing the same $i$ and $\phi$. Here, we again fix $r_o$ to 400 AU where beyond which the image fit does not get better and the SED fit does not show any significant improvement.
We first use least squares fitting to find the best-fit model parameters. This model is a much better fit comparing to the previous single continuous disk model. The total $\chi^2$ decreases by more than 3000. The contributions to the total chi-square from the broad band photometry, IRS spectrum, and images are 3\%, 14\%, and 83\%. The image fit improves significantly. The residual images do not suffer from the severe over subtraction in the central regions anymore. Therefore, it is worth exploring the uncertainties of the fitting parameters for this model. However, the covariance matrix from the least squares fitting does not always provide reasonable error estimations, especially for parameters that are degenerate. Degeneracies are prominent between \bem, $\Gamma$, $C$, and $r_i$ for the continuous disk model and between \bem, $C_r$, and $r$ for the ring model. For example, effects of temperature depression from increasing $r$ can be compensated by having a larger value of \bem. To better quantify the uncertainties on the parameters, we use the ensemble MCMC method of \citet{good10}, as implemented in the \texttt{emcee} python package \citep{fore13}.

The \texttt{emcee} package is based on the Markov chain Monte-Carlo (MCMC) method developed by \citet{good10} where they utilize multiple "walkers" to propagate multiple chains simultaneously. We set the initial positions of the walkers by drawing random numbers based on Gaussian distributions with means equal to the best-fit parameters and the variances equal to 10\% of the means. This MCMC algorithm adjusts the candidate step proposal distribution based on the walkers' current positions in the parameter space. We use 100 walkers and run for few thousand steps after the burn-in period seems to be over. All of the marginalized probability density functions (PDFs) look fairly symmetric. We calculate the uncertainty on each parameter by measuring the 1$\sigma$ interval on the marginalized PDFs, with the upper and lower bounds measured separately. The parameter values and their uncertainties are listed in Table~\ref{fitparams}.

The best fit shows that the continuous component contains hotter grains and is mainly contributing to the 11.7 \um\ emission while the ring component contains cooler grains and dominates the 18.3 \um\ emission. The hot continuous disk extends from a few tens of AU to hundreds of AU and contains grains as hot as 300 K. The cool ring locates around 60 AU, spatially overlapping with the continuous component. By comparing to the single-disk model, this two grain populations model significantly improved the fit to the images and to the short wavelength part of the SED. However, this model is still under predicting the flux longer than 35 \um\ on the SED. The main reason for causing this underprediction is the outnumbered \irs\ spectral data points. The fit is driven by closely matching the spectrum in order to effectively lower the $\chi^2$. As a result, the model completely misses the photometry points at wavelengths longer than 35 \um. As a test, we considered excluding long wavelength \irs\ points. We find that in order to produce a reasonably good fit with this two-component model, we have to cut out \irs\ points longer than $\sim$ 15 \um. Since we do not have a physical explanation to place such cut, we include all \irs\ points in our analysis. To account for the long wavelength emissions, we add another ring component in the following section.

\subsubsection{A Continuous Power-law Disk + Two Rings}
\label{model3}
\begin{figure}
\begin{center}
\hspace*{-0.15in}
\includegraphics[width=3.8in]{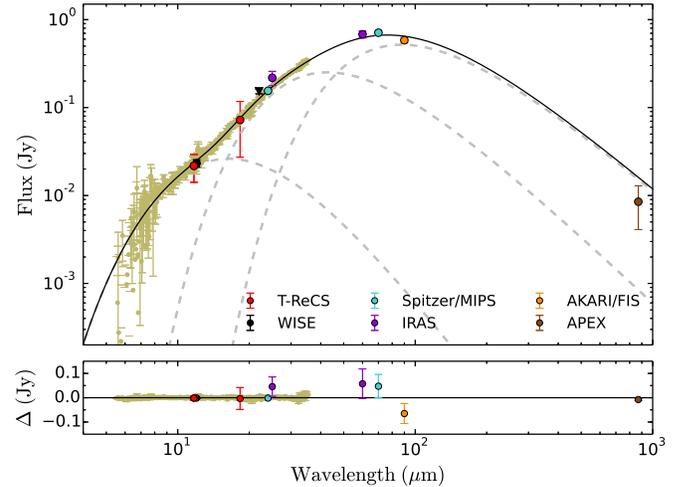}
\end{center}
\setlength{\abovecaptionskip}{-10pt}
\caption{SED of the debris disk around \hdtar\ with the best-fit three-population model (solid line) and its components (dashed lines). The residuals are plotted in the bottom panel. This model is composed of a hot continuous power-law disk, a warm ring, and a cold ring. The grains are assumed to emit like modified blackbodies such that the emissivity is proportional to $\nu ^\beta$. The best-fit parameters are listed in Table~\ref{fitparams}. Given this grain emissivity law, this three-component model is the simplest model that provides a reasonable fit to the SED while fitting the images simultaneously.
\label{sedfig1}}
\end{figure}

\begin{figure*}
\begin{center}
\includegraphics[width=7.in]{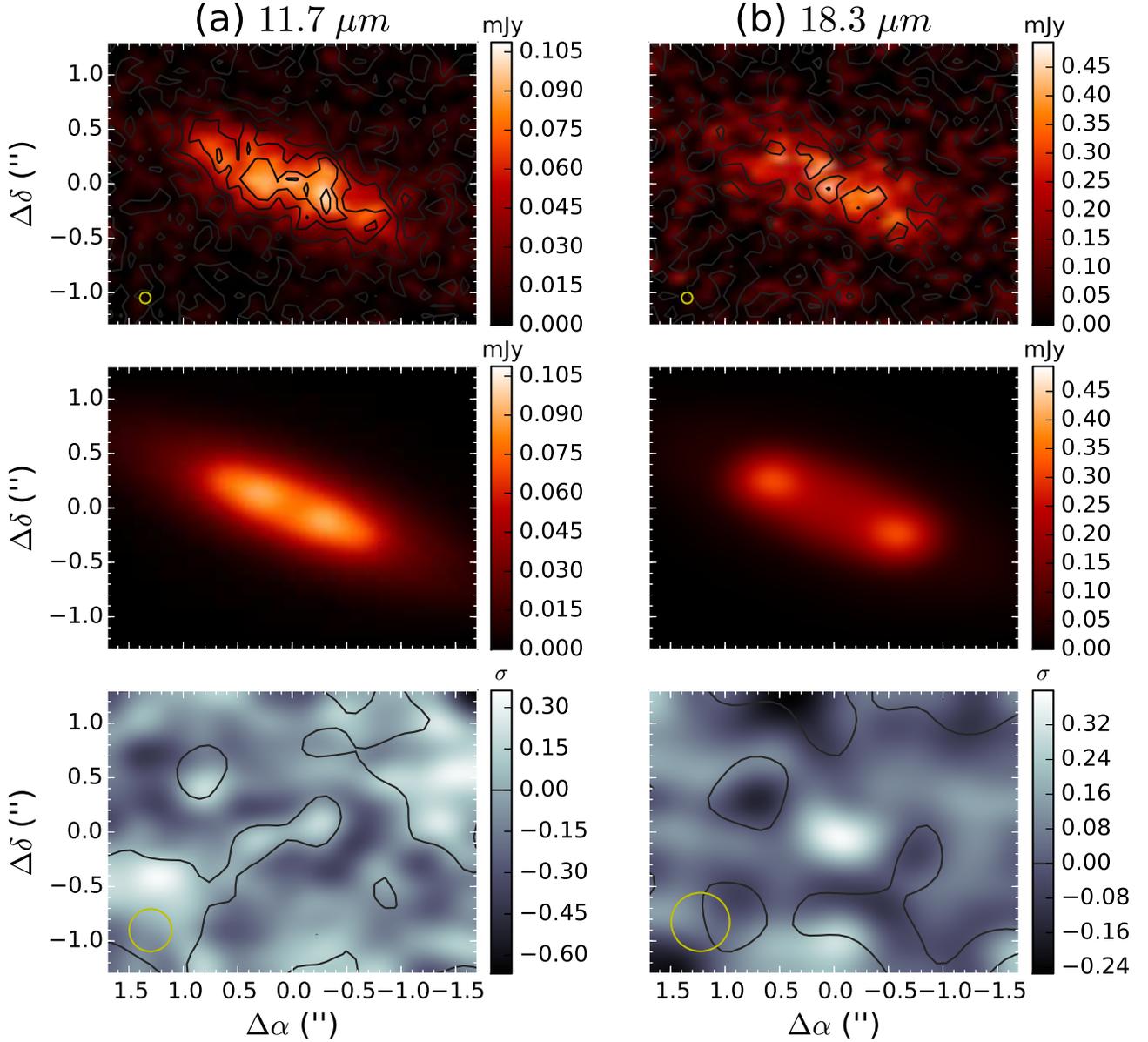}
\end{center}
\setlength{\abovecaptionskip}{-10pt}
\caption{Images showing the best fit of the three-component model (a continuous power-law disk + two rings). 
Top panel: the PSF subtracted images, smoothed to suppress high-spatial-frequency noise for the displaying purpose. We smoothed the images with a gaussian with its FWHM (yellow circle) equals to 0.1$\arcsec$. Contours are spaced by the 1-$\sigma$ background noise level derived from the smoothed variance map. 
Middle panel: the best-fit model convolved with the stellar PSF. 
Bottom panel: the smoothed residual flux divided by the uncertainty derived from the smoothed variance map. Images are smoothed with gaussians with FWHM equal to their PSF sizes only for the displaying purpose. 
Comparing large spatial structures in the residual images, this model seems to slightly under predict the central flux. Nonetheless, the variations in the resolution-element scale are within 1 $\sigma$, suggesting the model fits the images reasonably well. The best-fit parameters are listed in Table~\ref{fitparams}.
\label{imgfig1}}
\end{figure*}

This three grain populations model is composed of a continuous disk and two thin rings. The structures of these model components are described in the previous two sections. Although the inclination and the disk position angle are free parameters in the model, we constrain the three components to share the same values. As before, we use the least squares fitting to find the best-fit model parameters and MCMC to quantify their uncertainties. The model parameters are listed in Table~\ref{fitparams}. The contributions to the total chi-square from the broad band photometry, IRS spectrum, and images are 2\%, 12\%, and 86\%. Fig.~\ref{sedfig1} and Fig.~\ref{imgfig1} show the best-fit SED model and the image models respectively. 

This best-fit model is composed of a hot continuous power-law disk extended from $35\pm3$ AU to $310^{+30}_{-20}$ AU with temperature ranging from 330 K to 150 K, a warm ring with temperature of 97 K located at $105\pm5$ AU and the cold ring with temperature of 52 K located at $220\pm40$ AU. The middle panel of Fig.~\ref{imgfig1} shows the PSF convolved model disk in the two imaging wavelengths. Approximately 90\% of the flux in the 11.7 \um\ model comes from the hot continuous component while 10\% comes from the warm ring. At 18.3 \um, about 2/3 of the flux comes from the warm ring and 1/3 of the flux comes from the hot disk. This model provides a reasonable fit to the images, despite there are some small patches of over-subtraction and under-subtraction as seen from the image residuals (the bottom panel in Fig.~\ref{imgfig1}). We characterize the variations in the resolution-element scales by dividing the sum of the smoothed residual flux values in a resolution-element-sized area by the sum of the corresponding smoothed noise. The variations in the resolution-element scales are within 1 $\sigma$, suggesting the model fits the images reasonably well. The cold ring component is too faint to affect either imaging channel. However, this cold ring is essential for contributing to the long wavelength part of the SED and is a key reason why we can constrain $r_o$, unlike the previous two models. We found that this three-component model is be able to provide a sensible fit to the SED and the images simultaneously.

\subsubsection{Other Disk Models}
\label{model4}
We also tried using a broken-power-law disk model. This model is composed of one continuous-disk described by two power-law grain distributions: one between the inner radius and the intermediate radius and the other between the intermediate radius and the outer radius. The entire disk shares a single value of emissivity power index $\beta$, in which we assume the grain composition is the same across the disk. The best-fit result improves slightly compared to the one from a single continuous disk model described by only one power-law grain distribution (\S~\ref{model1}). However, this best-fit broken-power-law model suffers from the similar inconsistency as the single-power-law model where the model flux is too concentrated in the center of the disk in both bands. The reason lies under the assumption of a single grain population. Since the temperature of the entire disk is governed by a single emissivity power index, grains further away will have lower temperatures. Thus, although a steeply rising density power-law can make the model disk flux more extended, it will also drive the corresponding SED model too bright at longer wavelengths since the flux is originated from low-temperature grains. In order to have warm dust grains be responsible for the extended emission instead, we must introduce a second emissivity power index to our model (as used in \S~\ref{model2}).

Another model we tried is composed of two continuous disks. We found that the two components can share the same values for $\Gamma$, $r_i$, $r_o$, $i$, and $\phi$ while still producing a good fit compared to having two independent sets of parameters. However, this model is very sensitive to the $\beta$ values. A small adjustment in $\beta$ will introduce a significant deviation in the resulting model. Thus, we must keep the two $\beta$ parameters independent in order to maintain a good fit. This result again indicates there are more than one population of grains in the system. We notice that one of the continuous components can be simplified into a ring without changing the best-fit result significantly ($\chi^2_{\nu}$ changes from 0.65 to 0.64). Therefore, we performed the detailed analysis of a continuous power-law disk plus a ring model (\S~\ref{model2}) instead of the two-continuous component model. 


\section{DISCUSSION}
\label{discussion}

Our models support the inference of multi-spatial components from a multi-temperature SED. Recently, \citet{kenn14} argued that if the SED of a debris disk around an A-type star shows multiple temperature components, it is an indication that the system hosts multiple populations of grains in different locations. Our SED modeling of the disk around \hdtar\ indicates there are multiple temperatures. From modeling with the resolved images in mid-infrared, we confirm the system indeed have grains at multiple spatial locations: a hot continuous component extends from $35\pm3$ to $310^{+30}_{-20}$ AU, a warm ring located at $105\pm5$ AU, and a cold ring located at $220\pm40$ AU. Our model indicates the two separated narrow rings are embedded in an extended disk component. Although not all the model components are completely spatially separated, we are confident that the dust is not concentrated in a single belt. Our modeling result agrees with the argument made by \citet{kenn14}, adding to the small poll of high resolution observations confirmed cases.

Grains with effective temperatures hotter than blackbodies are responsible for the observed disk emission, since perfect blackbody grains at these spatial locations would not have the appropriate color temperatures. In our models, the grains are assumed to emit according to a modified blackbody function where the emissivity is proportional to $\nu^{\beta}$. In the case of the three-population model, the grains in the $\beta$ values for the continuous component, the warm ring, and the cold ring are $1.64\pm0.02$, $0.59\pm0.02$, and $0.32\pm0.06$ respectively. $\beta$  = 0 corresponds to blackbody grains. The positive $\beta$ values indicate the grains are small since they are inefficient in absorbing and emitting at long wavelengths, and $\beta$ = 1 corresponds to small grains in the limit $2 \pi a \ll \lambda$, where $a$ is the grain size and $\lambda$ is the observed wavelength. For $\beta$ > 1, real materials with complicated emissivities are required to explain the observations. Observational evidence points that it is common for debris disks around A-stars to have $\beta$ values approximately within 2 \citep{boot13}, and our results fall within this region.

\begin{figure}
\begin{center}
\hspace*{-0.22in}
\includegraphics[angle=0,width=3.94in]{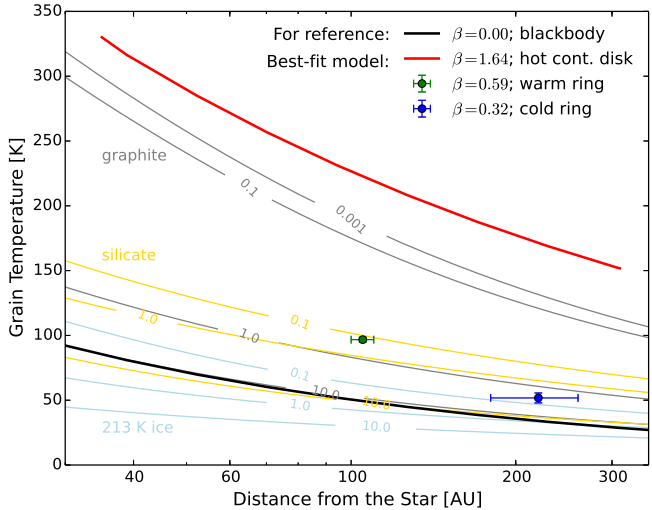}
\end{center}
\setlength{\abovecaptionskip}{-10pt}
\caption{Equilibrium temperatures of the disk model components in comparison to various grain types. The temperatures and radii for the three-component model are shown here. The red curve represents the hot continuous disk component. The green and blue dots represent the warm and cold ring components respectively. The black line is the temperature curve assuming grains are perfect blackbodies. Note that all three populations of grains have equilibrium temperatures higher than blackbodies. This grain property is commonly observed since small grains are inefficient in emitting at long wavelengths. The grain size (grain radii in \um) contours for graphite, silicate, and ice are overplotted here for comparison. Although the optical property of ice changes with temperature, we do not take it into account since the variation is quite small; here we use the T = 213 K ice crystal structure. Although small graphite grains, as expected, are significantly hotter than other grains, it still can not match the temperature gradient for the hot continuous disk component. Thus, grains composed solely of graphite, astro-silicate, or ice in the equilibrium temperature can not be used to explain all the observed disk properties.\label{Tg_r}}
\end{figure}

We searched for the potential grain compositions by first examining the equilibrium grain temperatures with our modified blackbody models and then compare them to the equilibrium temperatures of specific grain compositions. Using the model parameter $r$ and $\beta$, we plotted the corresponding equilibrium temperatures at different stellar locations in Fig.~\ref{Tg_r}. Our goal is to use these temperature curves to identify the possible grain compositions. We consider some common compositions such as graphite, astro-silicate, and ice. We use the optical constants of graphite and silicate from \citet{drai84} and \citet{laor93}. The optical property of ice is a function of its temperature due to varying crystal structures. \citet{warr84} provided the index of refraction of pure ice at $-1^{\circ}$C, $-5^{\circ}$C, $-20^{\circ}$C, and $-60^{\circ}$C. Although the optical properties are different at different crystal temperatures, the variation is quite small. Thus, to first order, we adopt the optical constants of ice assuming the ice preserves the similar crystal structure as when it is at $T = -60^{\circ}$C. We calculated their equilibrium temperatures using the Mie theory \citep{bohr83} with the assumption that grains are spherical. The grain size contours are overplotted on the temperature-stellocentric distance map on Fig.~\ref{Tg_r}.

Grains composed solely of graphite, astro-silicate, or ice in the equilibrium temperature can not be used to explain all the observed disk properties. Among the three compositions, graphite grains have the highest temperature, and ice grains are the coolest. We have also considered using the other types of silicates such as olivine and pyroxene. However, their optical properties are similar to the astro-silicate, and thus, the grain temperatures are on the same order of magnitude as the astro-silicate grains. The warm and cold rings could be made up by graphite and silicate grains closer to micron size. On the other hand, the hot continuous disk component would require the emitters to be really small. Since small grains become inefficient emitters at long wavelengths, these grains will sustain much higher temperatures than a blackbody. However, even though nanometer-sized graphite grains are significantly hotter than blackbodies, they still could not match the temperature of the hot continuous disk. In addition, grains this small are not likely to be present in the system as discussed below. 

We have qualitatively considered having porous grains and using a broken emissivity power-law in attempt to address the abnormally warm dust. However, neither considerations are in our favor. Porous dust grains have lower temperatures than compact spheres \citep{kirc13}. Thus, introducing porosity will make the discrepancy worse for finding a physical grain composition for the observed dust temperatures. Similar qualitative result applies for using a broken emissivity power-law case. A broken emissivity power-law assumes that the emissivity is 1 when $\nu > \nu_c$ and $(\nu_c/\nu)^{\beta}$ when $\nu < \nu_c$, where $\nu_c$ is a free parameter indicating the critical frequency at which the emissivity function changes. Compared to the smooth emissivity law we used, the grains with the broken emissivity power-law are less efficient in absorbing the stellar light. Therefore, grains with a broken emissivity power-law will sustain lower temperatures. In addition to making the grains cooler, a broken emissivity power-law introduces an additional model parameter, $\nu_c$, that makes the model more complex than a smooth emissivity power-law.  

Although the high disk-temperature draws the connection to small grains, no features of small grains are detected in the system. Nanometer-sized grains are in the molecular region. For \hdtar, we do not observe any solid-state features in the 10 and 20 micron regions from the \irs\ spectrum, which indicates that the grains are probably larger and more likely to be in the micrometer size region (e.g., \citealt{drai84}). Stochastically heating small grains such as polycyclic aromatic hydrocarbons (PAHs) could be one way to make grains have much higher temperatures than in the equilibrium states \citep{drai01}. However, the IRS spectrum does not show the spectral signatures of PAHs either.

Furthermore, small grains are unlikely to present in the system when considering the radiation pressure from the star. The blowout grain size is calculated by balancing the radiation pressure with the gravity. Taking the mass of the star to be 1.9 M$_{\sun}$ \citep{chen12} and using the grain density of $\rho = 3.5\ g/cm^3$, we get the blowout grain radius for \hdtar\ to be:
\begin{equation}
a_{blowout} = 0.91\ \mu {\rm m} \left(\frac{L_{\ast}}{10.5\ L_{\sun}}\right)\left(\frac{1.9\ M_{\sun}}{M_{\ast}}\right)\left(\frac{3.5\ g/cm^3}{\rho}\right).
\end{equation}
Grains smaller than this blowout size are unlikely to survive in the system. Although small grains can be trapped in resonance due to gas drag while migrating, the effect is unlikely to be significant with the gas level present in this system. Since small grains are unlikely to be responsible for the hot continuous disk emission, the nature of these abnormally warm grains is not completely understood.  

Identifying unique disk models around A stars can be challenging. For example, recently, with \herschel\ observations in 70, 100, and 160 \um, \citet{boot13} shows that the disks around A stars have various morphologies, ranging from systems that can be fit with just a narrow ring to the ones that require wider or multiple rings. Although direct images can provide constraints on grain size distribution and dust properties, sometimes finding a proper model can be quite difficult. For example, a detailed modeling work on $\beta$ Leo debris disc with multi-wavelength observations shows the degeneracy between one, two, and three components models \citep{chur11}. For \hdtar, the three-component model gives a reasonable fit but more data are needed to set better constraints. We would like to better characterize the grain properties, understand the distribution of grains, and constrain the location of planetesimal belts in much greater detail.

Fortunately, with the new generations of high resolution, high sensitivity, and high contrast instruments, detailed disk characterization is foreseeable. \hdtar\ is located in the Southern sky, making it a favorable target for GPI (\gpi), SPHERE (\sphere), and ALMA (\alma). GPI and SPHERE can potentially image the disk in scattered light in the near infrared, providing the currently uncharacterized scattering properties of the dust grains around \hdtar, in both total intensity and linearly polarized light. Designed specifically for high contrast imaging, these instruments have the potential to probe dust-scattered light at small inner working angles (e.g., \citealt{perr14}). With superior sensitivity and resolution, ALMA is capable for detailed mapping of the CO gas in this system. In addition, ALMA observations would trace the distribution of larger grains which are more tightly coupled to the locations of the large parent bodies. Future observations with these facilities hold great promise in further characterizing the dust distribution and dynamics.  

\section{SUMMARY}
\label{summary}

\hdtar\ shows strong infrared excess, and here we present the discovery of the resolved debris disk in the mid-infrared. The debris disk's properties can be constrained using all the available observations on \hdtar, including 15 photometry points, the \irs\ spectrum, and resolved images at 11.7 \um\ and 18.3 \um. From our modeling result, the disk clearly can not be described by a single continuous population of modified blackbody grains. The images alone can be described by a two grain population model which is composed of a continuous power-law disk and a ring. The continuous component contains hotter grains and dominates the emission in the 11.7 \um\ image whereas the ring component contains cooler grains and dominates the emission at 18.3 \um\ image. However, in order to obtain a good fit to the SED simultaneously, an additional ring component is needed. This three-component model is composed of a hot continuous power-law disk, a warm ring, and a cold ring. In this model, the disk fluxes in the imaged wavelengths are contributed primarily from the hot continuous disk and the warm ring. Since the cold ring peaks at a longer wavelength, this third component does not show up in the mid-infrared images so its spatial location is relatively unconstrained. The excess emission in far infrared and submillimeter, however, can be well described by this third component. Starting with the simplest model, we found that a model with three components and therefore three grain populations can well describe the images and the SED simultaneously.
\\

\section*{ACKNOWLEDGEMENTS}

Work by L.-W. Hung is supported by the National Science Foundation Graduate Research Fellowship number 2011116466 under Grant number DGE-1144087. P. G. Kalas and J. R. Graham thank support from NASA NNX11AD21G, NSF AST-0909188, and University of California LFRP-118057.

\bibliographystyle{apj}

\end{document}